\documentclass[12pt,epsfig,a4]{article} 
\newcommand{\vs}{\vspace{-0.25cm}}
\usepackage{amsmath,graphicx,wasysym}
\begin{document} 
\begin{center}
{\Large{\bf Radiative corrections to the magnetic moments 
of the proton and the neutron}\footnote{This work 
has been supported in part by DFG and NSFC (CRC110).}  }  

\bigskip

 N. Kaiser \\
\medskip
{\small Physik-Department, Technische Universit\"{a}t M\"{u}nchen,
   D-85747 Garching, Germany\\

\smallskip

{\it email: nkaiser@ph.tum.de}}
\end{center}
\medskip
\begin{abstract}
We estimate the radiative corrections of order $\alpha/\pi$ to the magnetic 
moments of the proton and the neutron. The photon-loop diagram of the 
vertex-correction type is evaluated with phenomenological nucleon vector form 
factors. Infrared-finiteness and gauge-invariance require the inclusion of the 
wave-function renormalization factor from the self-energy diagram. Using recent 
empirical form factor parametrizations the corrections amount to 
$\delta\kappa_p= -3.42 \cdot 10^{-3}$ and $\delta\kappa_n=
1.34 \cdot 10^{-3}$. We study also the effects from photon-loops with internal 
$\Delta(1232)$-isobars. For two customary versions of the $\Delta N\gamma
$-vertex and spin-3/2 propagator, these radiative corrections 
have values of $\delta\kappa_p^{(\Delta)}= (-0.9,\, 0.0)\!\cdot\! 10^{-3}$ and 
$\delta\kappa_n^{(\Delta)} = (1.2,\,-0.8)\!\cdot\! 10^{-3}$, respectively.
    
\end{abstract}

\section{Introduction}
The magnetic moments of the nucleon (proton and neutron) are important 
structure constants and their understanding represents a challenge for any 
model or theory of the strong interactions. The measured values of the proton 
magnetic moment $\mu_p =1+\kappa_p = 2.792847356(23)$ and the neutron magnetic 
moment $\mu_n =0+ \kappa_n = -1.9130427(5)$, both given in units of the nuclear 
magneton $e\hbar/2M_p$, are extremely precise \cite{pdg14}. The large anomalous 
magnetic moments $\kappa_p$ and $\kappa_n$, describing the deviations 
from the values $1$ and $0$ of a charged and neutral Dirac-particle, arise 
predominantely from the strong interactions in these composite three-quark 
systems. However, at the given experimental precision of $\kappa_p$ and 
$\kappa_n$, their full-fledged values cannot be attributed to the strong 
interactions alone, and electromagnetic effects (or radiative corrections) 
play  also a substantial role. In the case of charged leptons a leading 
correction of $\delta\kappa_\ell = \alpha/2\pi =  1.1614 \cdot 10^{-3}$ is 
provided by the one-photon loop diagrams, calculated first by J. Schwinger in 
1948. Due to the ubiquitous prefactor $\alpha/\pi$, one can expect that 
radiative corrections to the nucleon magnetic moments will similarly amount in 
magnitude to a few per mille. However, when deriving the analogous radiative 
corrections for nucleons one has to take into account that these composite 
objects possess an electromagnetic structure, preformed by the strong 
interaction, and that they are excitable into baryonic resonances. The purpose 
of this paper is to provide quantitative estimates for the radiative 
corrections to the nucleon magnetic moments based on parametrized nucleon form 
factors and a Lorentz-covariant treatment of the low-lying spin-3/2 
$\Delta(1232)$-resonance.

The present paper is organized as follows. In section 3 we calculate the 
one-photon loop diagrams of the vertex-correction and self-energy type with 
inclusion of phenomenological nucleon vector form factors. As prepared in 
section 2, the radiative correction to the anomalous magnetic moment is 
extracted by a projection technique and after Wick-rotation to euclidean 
momentum-space and angular integration one obtains a set of weight-functions. 
These serve to express 
the quantity of interest, $\delta \kappa = F_2^{(\gamma)}(0)$, as a one-parameter 
integral over a quadratic form in the nucleon vector form factors. Moreover, we 
verify that the inclusion of the wave-function renormalization factor $Z_2$ 
from the (electromagnetic) self-energy diagram leads to an infrared-finite and 
gauge-invariant result. Several parametrizations of the proton and neutron 
vector form factors are used to evaluate the integral-represention of $\delta 
\kappa_p$ and  $\delta \kappa_n$. Section 4 is devoted to the calculation of 
the photon-loop diagrams with a single intermediate $\Delta(1232)$-isobar. 
We employ common forms of the Rarita-Schwinger propagator and the $\Delta 
N\gamma$-vertex amended by a phenomenological transition form factor. Section 5 
deals with the mechanism, where a charged $\Delta^+$-isobar propagates further 
inside the photon-loop after electromagnetic interaction, which is 
applicable only to the proton. The pertinent $\Delta^+\Delta^+\gamma$-vertex 
is obtained by gauging the kinetic term of the free Rarita-Schwinger 
Lagrangian. In section 6 we 
study alternative forms of the Rarita-Schwinger propagator and the $\Delta N
\gamma$-vertex, which incorporate the feature of projecting onto the spin-3/2 
degrees of freedom. The results obtained with both Lorentz-covariant 
treatments of the $\Delta(1232)$-isobar may provide a range for the size of
baryonic resonance contributions to the radiative corrections to the 
nucleon magnetic moments.  Finally, section 7 ends with a short summary and in 
the appendix expressions for the electromagnetic mass shifts are collected and 
discussed. 

\section{Preparation}
Let us start with recalling the form of the on-shell vector-vertex for a
spin-1/2 nucleon of (average) mass $M=939\,$MeV. Imposing current-conservation 
and the usual discrete symmetries, the transition matrix-element 
$N(p_1)+\gamma^*(q)\to N(p_2)$ reads:
\begin{equation} \Gamma^\mu = \gamma^\mu\, F_1^{(\gamma)}(-t)+{i\over 2M}
\, \sigma^{\mu\nu} q_\nu\, F_2^{(\gamma)}(-t)\,, \end{equation}
with the spin-tensor $\sigma^{\mu\nu}=i( \gamma^\mu \gamma^\nu-\gamma^\nu 
\gamma^\mu)/2$ and the squared momentum-transfer $t= q^2=(p_2-p_1)^2\leq0$. The 
superscript $(\gamma)$ on the Dirac and Pauli form factors $F_{1,2}^{(\gamma)}(-t)$
should indicate that $\Gamma^\mu$ arises from a photon-loop diagram. In order 
to extract the radiative correction to the magnetic moment, one has to
project out first the loop-induced Pauli form factor $F_2^{(\gamma)}(-t)$. This 
is achieved by multiplying $\Gamma^\mu$ with a suitable projection operator 
\cite{jeger} and taking the Dirac-trace:
\begin{equation}  F_2^{(\gamma)}(-t)= {M \over t(4M^2-t)} {\rm Tr} 
\Big\{\Gamma^\mu(\gamma\!\cdot\!p_1+M) \Big[M \gamma_\mu +{t+2M^2 \over 
t-4M^2} (p_1+p_2)_\mu \Big](\gamma\!\cdot\!p_2+M)\Big\} \,. 
\end{equation}
In the next step the limit $t\to 0$ has to be performed with care before the 
four-dimensional integration over the loop-momentum $l$. By choosing the 
Breit-frame with $q_0=0$ and averaging over the directions of the (small) 
momentum-transfer $\vec q$, one can effectively make the substitution: 
$(l\!\cdot\! q)^2 = (\vec l \!\cdot\! \vec q\,)^2 \to \vec l^{\,2}\vec q^{\,2}/3 
= -\vec l^{\,2}t/3$. During this procedure the momentum-sum $p_1+p_2= (\sqrt{4M^2
+\vec q^{\,2}},\vec 0)$ is  kept fixed. After setting $t=0$, the resulting 
loop-integrand depends only on 
$l^2$ and $l\!\cdot\!p_{1,2}= M l_0$.  One performs a Wick-rotation to euclidean 
momentum-space, $l_0 = i l_4$, and sets $l_4 = \ell z$ with $z=\cos \theta $ 
the cosine of the polar angle $\theta $ in four dimensions. By combining these 
steps, a loop-integral including the factor $-1/l^2$ from a photon-propagator 
gets reduced to:
\begin{equation} \int\!\! {d^4 l \over (2\pi)^4 i}\, {1\over(-l^2)}
\big[ \dots\big] = {M^2 \over 4\pi^3} \int_0^\infty \!\!dx\int_{-1}^1\!\!
dz\,x\, \sqrt{1-z^2}\,\big[ \dots\big]\,, \end{equation}
where one has set the euclidean length to $\ell= x M$. The remaining terms 
$[\dots]$ of the loop-integrand depend rationally on $z$, such that the 
angular integral can always be performed analytically. The outcome of the 
loop-calculation at this stage are $x$-dependent weight-functions. 

In the following sections we apply these techniques to calculate the radiative 
corrections to the nucleon magnetic moments as they arise from photon-loop 
diagrams involving nucleons and $\Delta(1232)$-isobars. The pertinent 
electromagnetic vertices are supplemented by phenomenological form factors which
should represent the (internal) baryon structure.      
\section{Photon-loop diagrams with nucleons}
\begin{figure}[t!]
\begin{center}
\includegraphics[scale=0.6,clip]{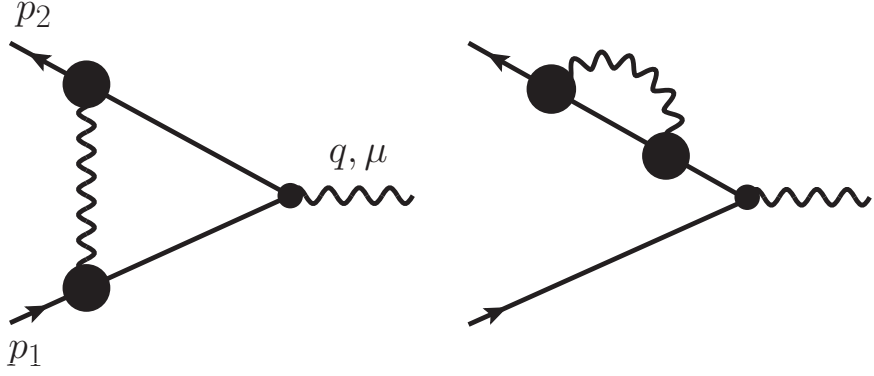}
\end{center}
\vspace{-.6cm}
\caption{Left: Triangular photon-loop diagram with intermediate 
nucleons. Right: Electromagnetic self-energy correction on an external 
nucleon line. The corresponding mirror diagram is not shown.}
\end{figure}
The triangular photon-loop diagram with intermediate nucleons is shown on the 
left in Fig.\,1. We evaluate it with three (off-shell) vector-vertices of the 
form $\Gamma^\mu$ in eq.(1), where the form factors  $F_{1,2}(\ell)$ and 
$F_{1,2}(0)$ refer to the nucleon structure arising from strong interactions.
Carrying out the calculational steps as outlined in section 2, one arrives at 
the following (partial) result for the radiative correction to the nucleon 
magnetic moments: 
\begin{eqnarray} F_2^{(\gamma)}(0) &\!\!\! =\!\!\! & {\alpha \over \pi} 
\int_0^\infty\!\!dx\, \bigg\{\bigg[\bigg({2+4x^2+x^4\over \sqrt{4+x^2}}
-2x-x^3\bigg)c+ \bigg({3+x^2\over \sqrt{4+x^2}}-x-{1\over x}\bigg)
\kappa\bigg]F_1^2(xM)\nonumber \\ && +\bigg[\bigg({x^2(7+2x^2)\over
\sqrt{4+x^2}}-3x-2x^3\bigg)c +\bigg(2x+3x^3-{x^2(8+3x^2)\over\sqrt{4
+x^2}}\bigg){\kappa \over 8}\bigg]F_1(xM)F_2(xM) \nonumber \\ &&
+ \bigg[\bigg({x^2\over \sqrt{4+x^2}}(16+x^2-x^4)-8x-3x^3+x^5\bigg)
{c \over 8} \nonumber \\ && + \bigg({x^2\over\sqrt{4+x^2}}(8+5x^2+x^4)
-4x-3x^3-x^5\bigg){\kappa \over 16} \bigg]F_2^2(xM)\bigg\}\,.
\end{eqnarray}
Here, $c=1,0$ denote the charge and $\kappa=1.793, -1.913$ the 
anomalous magnetic moment of the proton and neutron, respectively. Of course, 
in the end the third decimal place of $\kappa_{p,n}$ should be changed in 
order to treat properly only the strong interaction part. $\alpha= e^2/4\pi = 
1/137.036$ denotes the usual fine-structure constant. It should be noted that 
the triangle diagram has been evaluated with a photon-propagator 
$-i g^{\mu\nu}/l^2$ in Feynman gauge. The integral $\int_0^\infty\!\!dx$ over the 
first weight-function $(2+4x^2+x^4)/\sqrt{4+x^2}-2x-x^3$ has the value $1/2$, 
and this way one recovers the famous result $ F_2^{(\gamma)}(0)=\alpha/2\pi$ of 
Schwinger for a structureless spin-1/2 Dirac-particle.

The partial result in eq.(4) has an obvious problem in the form of the 
infrared-singular $1/x$-term in the weight-function multiplying $\kappa 
F_1^2(xM)$. The only other diagrams available at order $\alpha$ are those with
a self-energy correction on an external nucleon line, shown on the right in
Fig.\,1. Their effect is to multiply the tree-level vector-vertex with the 
nucleon wave-function renormalization factor $Z_2$. In order to extract this 
quantity, one decomposes the Dirac self-energy $\Sigma$ into a scalar and a 
vector part: $\Sigma =\Sigma_s(p^2)+\gamma\!\cdot\!p\,\Sigma_v(p^2)$. By setting
thereafter $p=(p_0,\vec 0 )$ the wave-function renormalization factor $Z_2$ is 
obtained as the derivative of $\Sigma_s(p_0^2) +p_0 \Sigma_v(p_0^2)$ with 
respect $p_0$ at $p_0=M$. The electromagnetic mass-shift follows in a simpler 
way as: $\delta M = \Sigma_s(M^2) +M \Sigma_v(M^2)$. After some calculation 
one arrives at the following expression for the wave-function renormalization 
factor $Z_2$ from the self-energy subdiagram in Fig.\,1:
\begin{eqnarray} Z_2 &\!\!\!=\!\!\!& {3\alpha \over 4\pi} \int_0^\infty
\!\!dx\,\bigg\{\bigg[{4\over 3x}-x^3+{x^4+2x^2-4 \over \sqrt{4+x^2}}
\bigg] F_1^2(xM)\nonumber \\ &&+\bigg[{x^2(16+5x^2)\over 2\sqrt{4+x^2}}
-3x-{5x^3\over 2}\bigg]F_1(xM)F_2(xM) \nonumber \\ &&+ \bigg[{x^2\over 
8\sqrt{4+x^2}}(16+2x^2-x^4)-x-{x^3\over 2}+{x^5 \over 8}\bigg]F_2^2(xM)
\bigg\}\,.\end{eqnarray}
One observes that in the combination $F_2^{(\gamma)}(0)+Z_2 \kappa$ the singular
$1/x$-term drops out, and thus an infrared-finite radiative correction is 
achieved. Moreover, one has to check gauge-invariance by working with a general 
photon-propagator:
\begin{equation} {i \over -l^2} \bigg( g^{\mu\nu} + \xi\, {l^\mu l^\nu 
\over -l^2}\bigg) \,. \end{equation} 
Its $\xi$-dependent longitudinal part produces via the triangular photon-loop  
diagram a contribution: 
\begin{equation}  F_2^{(\gamma)}(0) =-{\alpha \xi \over 2\pi}\, \kappa 
\int_0^\infty\!\!dx\,x^{-1} F_1^2(xM)\,, \end{equation} 
which gets exactly canceled by a contribution to the $Z_2$-factor from the 
self-energy subdiagram:
\begin{equation}  Z_2 ={\alpha \xi \over 2\pi} 
\int_0^\infty\!\!dx\,x^{-1} F_1^2(xM)\,. \end{equation}
The infrared-finite and gauge-invariant total result $F_2^{(\gamma)}(0)+Z_2 
\kappa$ from the photon-loop diagrams of vertex-correction and self-energy 
type in Fig.\,1 reads finally:
\begin{eqnarray} F_2^{(\gamma)}(0) &\!\!\! =\!\!\! & {\alpha \over \pi} 
\int_0^\infty\!\!dx\, \bigg\{\bigg[\bigg({2+4x^2+x^4\over \sqrt{4+x^2}}
-2x-x^3\bigg)c+ \bigg({x^2(10+3x^2)\over \sqrt{4+x^2}}-4x-3x^3\bigg)
{\kappa\over 4}\bigg]F_1^2(xM)\nonumber \\ && +\bigg[\bigg({x^2(7+
2x^2)\over\sqrt{4+x^2}}-3x-2x^3\bigg)c +\bigg({x^2(10+3x^2)\over
\sqrt{4+x^2}}-4x-3x^3\bigg){\kappa \over 2}\bigg]F_1(xM)F_2(xM) 
\nonumber \\ && + \bigg[\bigg({x^2\over \sqrt{4+x^2}}(16+x^2-x^4)-8x-
3x^3+x^5\bigg){c \over 8} \nonumber \\ &&+\bigg({x^2\over\sqrt{4+x^2}}
(64+16x^2-x^4)-32x-18x^3+x^5\bigg){\kappa \over 32} \bigg]F_2^2(xM)
\bigg\}\,,\end{eqnarray} 
with $c=1, 0$ the charges and $\kappa=1.793,-1.913$ the anomalous magnetic 
moments of nucleons.

The numerical evaluation of the integral-formula in eq.(9) for the radiative 
correction to the nucleon magnetic moments requires as input the 
momentum-dependent form factors $F_{1,2}(Q)$. The simplest $Q$-dependence,
which roughly reproduces the trend of the data, is a dipole parametrization of 
the electric and magnetic form factor \cite{nucleon}:
\begin{equation} G_E(Q) = F_1(Q) -{Q^2\over 4M^2} F_2(Q) = c\bigg(
1+{Q^2 \over \Lambda^2}\bigg)^{-2}\,, \end{equation}
\begin{equation}G_M(Q) = F_1(Q) +F_2(Q) = (c+\kappa) \bigg(1+{Q^2 
\over \Lambda^2}\bigg)^{-2}\,, \end{equation}
with $\Lambda = 843\,$MeV the empirical dipole mass. More elaborate 
parametrizations of the nucleon form factors are obtained in the dispersion 
relation analysis (DRA) \cite{dispersion}. In addition to the data from elastic 
electron-scattering, fits based on this method include the constraints from 
unitarity, analyticity and crossing symmetry as well as the QCD-asymptotics.
We employ here the expressions for the isoscalar and isovector form factors 
$F_{1,2}^{(s,v)}(Q)$ given in appendix C of ref.\cite{dispersion}. Moreover, 
improved fits of the nucleon vector form factors have been obtained recently in 
refs.\cite{ina1,ina2,ina3} by including two-photon exchange corrections in the 
analysis of all available and updated elastic electron-proton scattering data. 
The fitted results for $F_{1,2}^{(s,v)}(Q)$ are presented in an analytical form 
as sums of vector-boson poles supplemented by rational parametrizations of the 
contributions from the isovectorial $\pi\pi$-continuum \cite{ina4}.  

In Table\,1 numerical values for the radiative correction
$\delta \kappa_{p,n} = F_2^{(\gamma)}(0)$ to the proton/neutron magnetic moment
are listed. The three columns refer to the dipole parametrization (dipole), 
the earlier fit \cite{dispersion} using dispersion relations (DRA), and the 
most recent fit \cite{ina4} including $2\gamma$-exchange corrections (Ina). 
Modulo some slight variations, the radiative 
correction $\delta\kappa_p \simeq -3.4\cdot 10^{-3}$ to the proton magnetic 
moment is in magnitude about 2.5 times larger than $\delta\kappa_n\simeq 
1.3\cdot 10^{-3}$ for the neutron. In both cases the radiative corrections 
amount to about $-1\permil$ of the empirical magnetic moments $1+\kappa_p = 
2.793$ and $\kappa_n = -1.913$. In order to illustrate how the numerical 
values in Table\,1 have come about, we present in Fig.\,2 the integrand for 
$\delta\kappa_p$, without the small prefactor $\alpha/\pi$. One observes 
distributions that take on negative values for $x>0.05$ and peak around 
$x= 0.3$, corresponding to a momentum-transfer of $Q = \Lambda/3$ (with 
$\Lambda=843$\,MeV the dipole mass). The fast transition from positive to 
negative values has its origin in the six $x$-dependent weight-functions 
entering eq.(9), where all exept the first one are negative functions. The 
upper full curve in Fig.\,2 (with total area $1/2$ under 
it) shows the different situation for the radiative correction to the magnetic 
moment of a (structureless) lepton. Likewise, the integrand for $\delta
\kappa_n$ (without the factor $\alpha/\pi$) is displayed in Fig.\,3. In this 
case essentially only the term in the last line of eq.(9) contributes. Since 
a negative weight-function gets multiplied by $\kappa_n = -1.913$, one obtains 
a positive integrand and a positive radiative correction  of 
$\delta\kappa_n\simeq 1.3\cdot 10^{-3}$. 

At this point one can draw the interesting conclusion that the large anomalous 
magnetic moments $\kappa_{p,n}$ of nucleons, predetermined by the strong 
interaction, change the sign of the leading radiative correction proportional 
to $\alpha/\pi$ in comparison to pointlike leptons.      

 \begin{table}[t!]
\begin{center}
\begin{tabular}{|c|ccc|}
    \hline
form factor & dipole & DRA & Ina \\ \hline  $10^3\cdot\delta\kappa_p$ 
& $-$3.466  & $-$3.495 &$-$3.423  \\ $10^3\cdot\delta\kappa_n$ 
& 1.370  &  1.336 &1.337  \\ 
  \hline
  \end{tabular}
\end{center}
{\it Tab.1: Numerical values of radiative corrections to the proton 
and neutron magnetic moments.}
\end{table}

\begin{figure}[t!]
\begin{center}
\includegraphics[scale=0.5,clip]{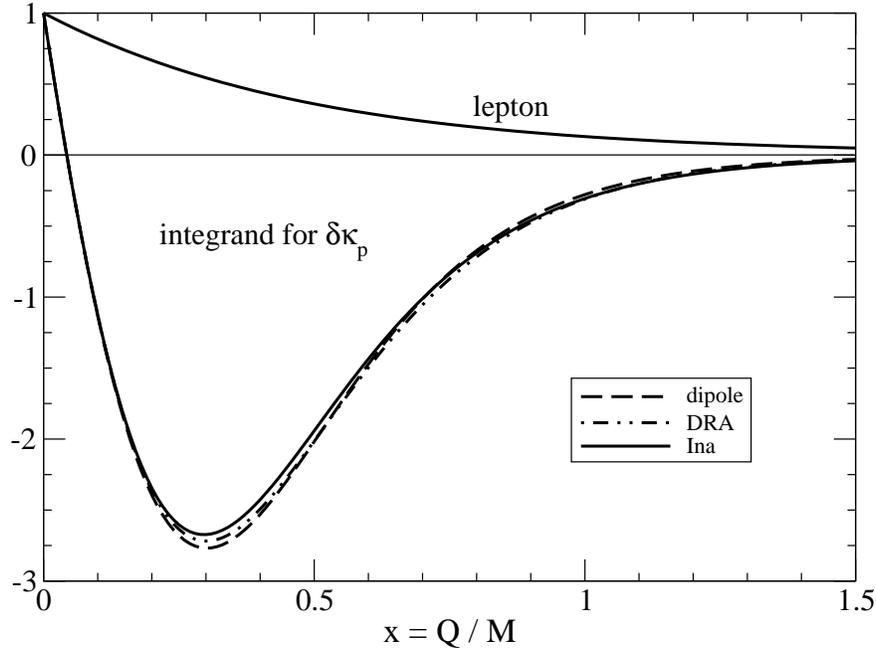}
\end{center}
\vspace{-.6cm}
\caption{Integrand (without factor $\alpha/\pi$) for the radiative 
correction to the proton magnetic moment.}
\end{figure}

\begin{figure}[h!]
\begin{center}
\includegraphics[scale=0.5,clip]{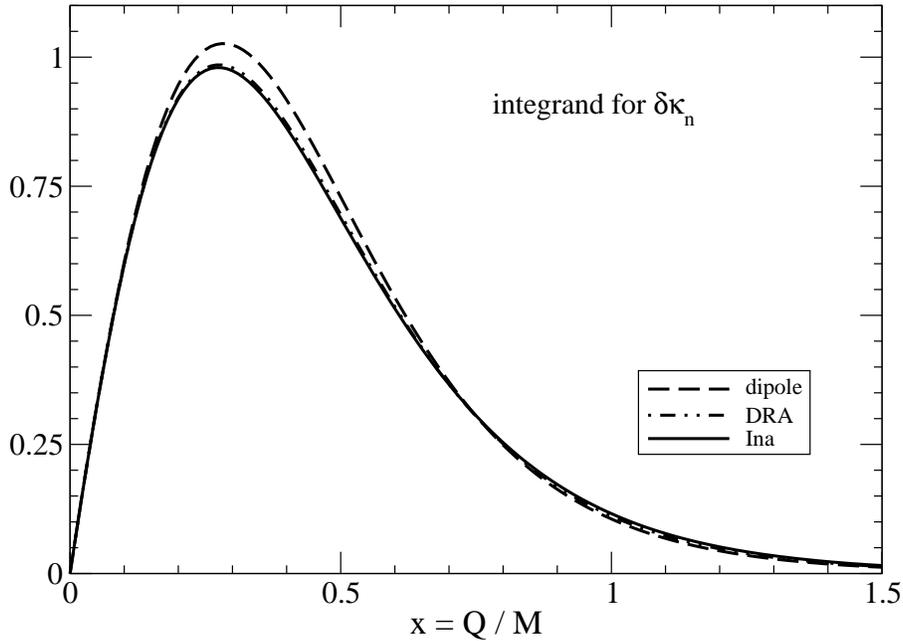}
\end{center}
\vspace{-.6cm}
\caption{Integrand (without factor $\alpha/\pi$) for the radiative 
correction to the neutron magnetic moment.}
\end{figure}

\section{Photon-loop diagrams with nucleon and delta-isobar}
In the following sections we study effects of the $\Delta(1232)
$-resonance on radiative corrections to the nucleon magnetic moments. The
$\Delta(1232)$-isobar with isospin-3/2 and spin-3/2 is the lowest-lying and 
most prominent excitation of the nucleon and the charge-states $(\Delta^+,
\Delta^0)$ decay also electromagnetically 
into $(p\gamma,n\gamma)$ final-states. The pertinent electromagnetic transition 
matrix-elements are related to measurable helicity amplitudes \cite{nucleon}. 
The present calculation of photon-loop diagrams requires yet a 
Lorentz-covariant formulation in which the spin-3/2 delta-isobar is described 
by a Rarita-Schwinger spinor $\Psi_\alpha$. There exists an extensive 
literature \cite{benmer,kondr,vanderh} about phenomenological applications as 
well as  potential conceptual problems of the Rarita-Schwinger formalism 
employed for the 
$\Delta(1232)$-resonance. A commonly used (minimal) form of the 
$\Delta N\gamma$-vertex reads \cite{benmer,bkmreview}:
\begin{equation} V^{\mu \alpha}_1 = {ie\kappa^*\over\sqrt{6}M} 
\big(g^{\mu \alpha} \gamma\!\cdot\!l -\gamma^\mu l^\alpha\big)\gamma_5\,
G_\Delta(\sqrt{-l^2})\,,\end{equation}
with $\mu$ the vector-index to couple the virtual photon with incoming 
four-momentum $l$. The stated vertex includes (in the prefactor) an isospin 
Clebsch-Gordan coefficient $\langle{3\over2}{\pm 1\over2}|10,{1\over2}{\pm 
1\over2}\rangle= \sqrt{2/3}$ and a phenomenological transition form factor 
$G_\Delta(\sqrt{-l^2})$, normalized to $G_\Delta(0)=1$. For the $\Delta\to N\gamma
$ transition magnetic moment $\kappa^*$  we choose the large-$N_c$ value 
$\kappa^* = 3\sqrt{2}(1+\kappa_p-\kappa_n)/4 \simeq 5.0$, which agrees well 
with empirical determinations \cite{benmer}. Note that the off-shell vertex in 
eq.(12) is gauge-invariant be construction, $l_\mu V^{\mu \alpha}_1 = 0$, and 
therefore the $\xi$-dependent part of the photon-propagator in eq.(6) will drop 
out immediately in our loop-calculation. Moreover, it should be mentioned that 
extensions of the vertex $V^{\mu \alpha}_1$ with further coupling constants and 
off-shell parameters have 
been proposed \cite{benmer,bkmreview}, but these parameters are not well 
determined. Another relevant ingredient for our loop-calculation is the  
Rarita-Schwinger propagator, whose common form reads \cite{benmer,bkmreview}: 
\begin{equation}{i\over 3}\, {\gamma\!\cdot\!P +M_\Delta \over 
M_\Delta^2-P^2}\bigg(3 g_{\alpha\beta} -\gamma_\alpha \gamma_\beta-{2P_\alpha 
P_\beta \over M_\Delta^2} +{P_\alpha \gamma_\beta-\gamma_\alpha 
P_\beta\over M_\Delta} \bigg)\,,\end{equation} 
with $P$ the four-momentum of the delta-isobar propagating from index $\beta$ 
to index $\alpha$.
 
\begin{figure}[t!]
\begin{center}
\includegraphics[scale=0.6,clip]{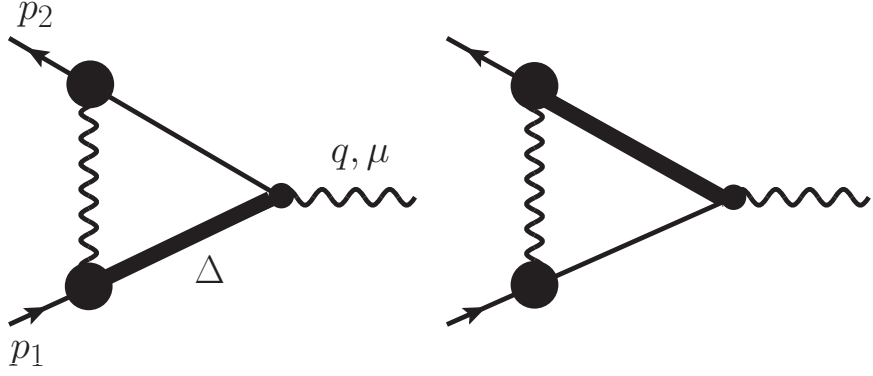}
\end{center}
\vspace{-.6cm}
\caption{Triangular photon-loop diagrams with intermediate 
nucleon und delta-isobar. Both diagrams contribute equally.}
\end{figure}
Fig.\,4 shows the two triangular photon-loop diagrams where one internal 
nucleon is excited (electromagnetically) into a $\Delta(1232)$-isobar. Since both 
diagrams are mirror partners of each other they contribute equally to the 
on-shell vector-vertex $\Gamma^\mu$. Carrying out the necessary calculations, 
one arrives at the following contribution from the photon-loop diagrams in 
Fig.\,4 to the radiatively induced magnetic moments of nucleons: 
\begin{eqnarray} F_2^{(\gamma)}(0) &\!\!\! =\!\!\! & {\alpha \kappa^{*2}
\over 54\pi r^2 } \int_0^\infty\!\!dx\, G_\Delta(xM)\bigg\{F_1(xM)
\bigg[{2\over x}(r^2-1)^2(r^2-r-2r^3)-6x(r+r^3+2r^5)\nonumber \\ && 
-6x^3(2+r+r^2+2r^3)-x^5(2+r)+{x^2\over r^2-1}\sqrt{4+x^2}\Big(4r^2-8-6r-
6r^3\nonumber \\ && -x^2(2+3r+2r^2+3r^3)\Big) +2r \sqrt{(r^2-1+x^2)^2
+4x^2}\bigg({r^2-1\over x}(1-r+2r^2)\nonumber \\ && +{x\over r^2-1}
\Big(2+r+r^3+4r^4+(1+2r+2r^2)x^2\Big)\bigg)\bigg]\nonumber \\ && +
F_2(xM) \bigg[-{r\over 2x}(r^2-1)^2(1+4r+3r^2)-x (r+9r^2+4r^3+3r^4+7r^5)
\nonumber \\ && -3x^3r^2 (2+3r)+{x^5\over 2}(6+r+x^2)
+{x^2\over r^2-1}\sqrt{4+x^2}\bigg(-8r^2(1+r)\nonumber \\ && 
+{x^2\over 2}\Big(4+r-14r^2-7r^3+(1+r-r^2)x^2\Big)\bigg)+{r\over 2}
\sqrt{(r^2-1+x^2)^2+4x^2}\bigg({r^2-1\over x}\nonumber \\ &&\times
(1+4r+3r^2)  +{x\over r-1}(1+13r-9r^2+11r^3)+{x^3\over r^2-1}
(7r^2+10r-1-x^2)\bigg)\bigg]\bigg\}\,, \nonumber \\ \end{eqnarray} 
with $r= M_\Delta/M = 1.312$ the ratio between the delta mass and the nucleon 
mass. One should note that the (lengthy) weight-functions multiplying 
$F_{1,2}(x M)$ behave near $x=0$ as order $x^2$. In order to evaluate the 
integral in eq.(14) numerically we use for the transition form factor 
$G_\Delta(xM)$ a dipole multiplied by an exponential function:
\begin{equation}G_\Delta(Q) = \bigg(1+{Q^2 \over \Lambda^2}\bigg)^{-2}
\exp\bigg(-{Q^2\over 7\Lambda^2}\bigg)\,. \end{equation}
Such an approximate $Q$-dependence has been extracted in ref.\cite{burkert} 
from pion-electroproduction data in the $\Delta(1232)$-resonance region.
 \begin{table}[t!]
\begin{center}
\begin{tabular}{|c|ccc|}
    \hline
form factor & dipole & DRA & Ina \\ \hline  $10^3\cdot\delta\kappa_p$ 
& $-$2.705  & $-$2.750 &$-$2.695  \\ $10^3\cdot\delta\kappa_n$ 
& 2.367  &  2.396 &2.412  \\ 
  \hline
  \end{tabular}
\end{center}
{\it Tab.2: Radiative corrections to the proton and neutron magnetic moments 
arising from $\Delta N\gamma$ triangle-diagrams.}
\end{table}

Table 2 gives the numerical values for $\delta\kappa_p$ and $\delta\kappa_p$
from the $\Delta N\gamma$ loop-diagrams for the three choices (dipole, DRA, Ina)
of the nucleon vector form factors $F_{1,2}(xM)$. One obtains sizeable 
contributions of $\delta\kappa_p\simeq -2.7 \cdot 10^{-3}$ and $\delta\kappa_n
\simeq 2.4 \cdot 10^{-3}$ with the same sign as those from the $NN\gamma$ 
loop-diagrams. The corresponding $x$-dependent integrands (without the factor 
$\alpha/\pi$) are shown by the full lines in Fig.\,5. Due to the suppression 
at low momentum-transfer, which stems from the extra-factor $l^\alpha$ in the 
vertex $V^{\mu \alpha}_1$, these distributions have peaks that are shifted upward 
to $x\simeq 0.4$.      

\begin{figure}[h!]
\begin{center}
\includegraphics[scale=0.5,clip]{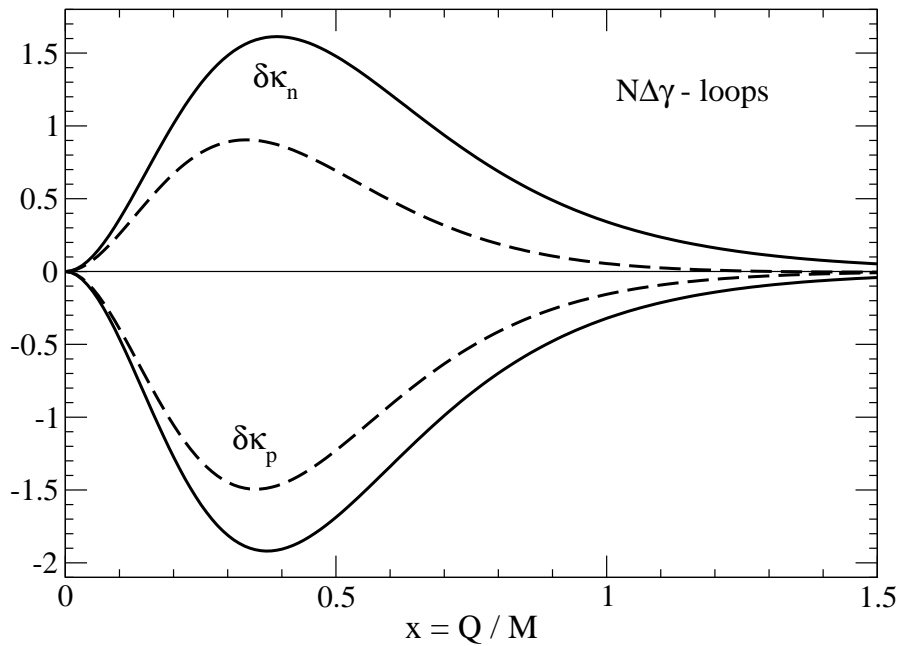}
\end{center}
\vspace{-.6cm}
\caption{Integrand (without factor $\alpha/\pi$) for the radiative 
corrections from $\Delta N\gamma$-loops.}
\end{figure}
\begin{figure}[t!]
\begin{center}
\includegraphics[scale=0.6,clip]{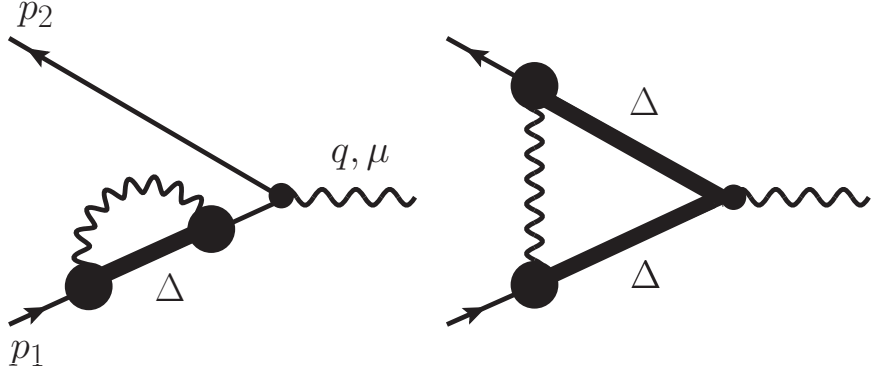}
\end{center}
\vspace{-.6cm}
\caption{Left: Electromagnetic self-energy correction with 
virtual delta-isobar excitation on a nucleon line. Right: Photon-loop 
diagram with two intermediate (charged) delta-isobars.}
\end{figure}
The next type for radiative correction that has to be considered is shown on 
the left of Fig.\,6. There, an electromagnetic self-energy insertion with an 
intermediate $\Delta(1232)$-isobar excitation is attached to the external 
nucleon lines. The nucleon wave-function renormalization factor $Z_2^{(\Delta)}$ 
derived from the self-energy subdiagram including a $\Delta(1232)$-isobar 
is given by the expression:
\begin{eqnarray}Z_2^{(\Delta)}&\!\!\! =\!\!\! & {\alpha\kappa^{*2}\over 
\pi(12r)^2} \int_0^\infty\!\!dx\, G_\Delta^2(xM)\bigg\{ {(r^2-1)^2
\over x}(5+10r^2+9r^4)+12x(1+2r^5+r^6)\nonumber \\ && +6x^3(3+4r^3)
+4x^5+3x^7 +\sqrt{(r^2-1+x^2)^2+4x^2}\bigg[{1-r^2\over x}(5+10r^2+9r^4)
\nonumber \\ &&+x(7-8r^2-24r^3-3r^4)+x^3(3r^2-1)-3x^5+{24r^3x(1+r)\over
(1+r)^2+x^2}\bigg]\bigg\}\,,
\end{eqnarray} 
and obviously it involves the square of the transition form factor 
$G_\Delta(xM)$. The corresponding contribution to the nucleon magnetic moments is
$\delta \kappa = F_2^{(\gamma)}(0)=Z_2^{(\Delta)} \kappa$. Using the dipole times 
exponential form of $G_\Delta(xM)$, one finds radiative corrections to the 
magnetic moments with values of $\delta \kappa_p=1.139\cdot 10^{-3}$ and $\delta 
\kappa_n =-1.216\cdot 10^{-3}$. Note that these reduce the previous values from 
the $\Delta N\gamma$-loop to about one half in size. For further illustration 
we show by the full line  in Fig.\,7 the integrand of $Z_2^{(\Delta)}$ (again 
without the factor $\alpha/\pi$). The displayed distribution starts at small 
momentum-transfer as order $x^3$ and develops a peak around $x = 0.5$.  
\begin{figure}[h!]
\begin{center}
\includegraphics[scale=0.5,clip]{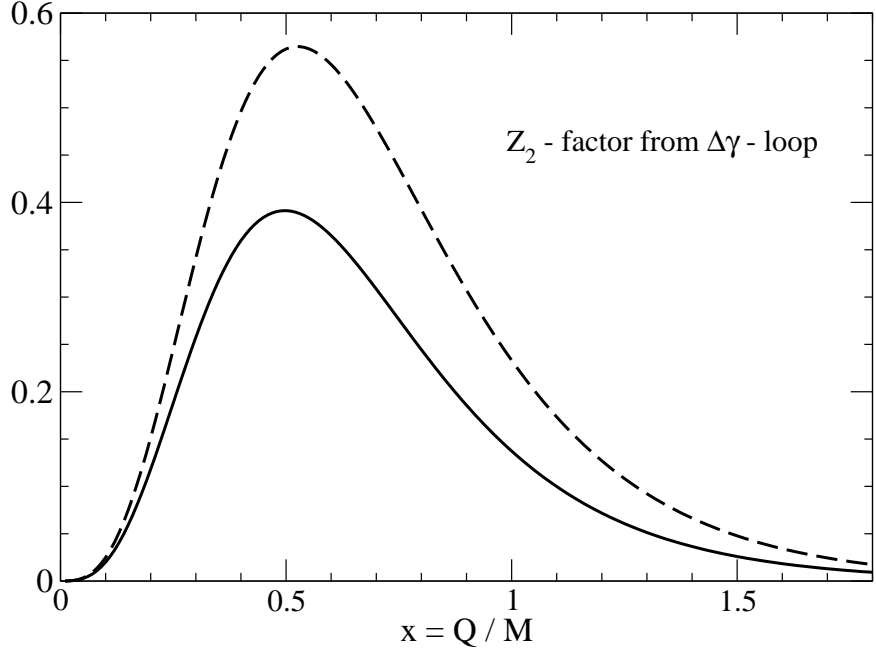}
\end{center}
\vspace{-.6cm}
\caption{Integrand (without factor $\alpha/\pi$) for the wave-function 
renormalization factor $Z_2^{(\Delta)}$ from the $\Delta\gamma$-loop.}
\end{figure}
\section{Photon-loop diagram with two charged delta-isobars}
Once the $\Delta(1232)$-isobar is introduced as an explicit degree of freedom, 
one can let it propagate further in the loop after coupling to the external 
photon. This mechanism requires a charged  $\Delta^+$ and thus it is applicable 
only to the proton. The relevant $\Delta^+\Delta^+\gamma$ coupling is obtained 
directly by gauging the kinetic term of the free Rarita-Schwinger Lagrangian 
\cite{benmer,bkmreview}, and it reads:
\begin{equation} i e\big( -\gamma^\mu g_{\alpha\beta} + \gamma_{\alpha} 
g_\beta^\mu + \gamma_{\beta} g_\alpha^\mu - \gamma_\alpha\gamma^\mu\gamma_\beta
\big)\,, \end{equation} 
with $\beta$  the incoming and $\alpha$ the outging (Rarita-Schwinger) index.
Note that no form factor is necessary here, since this vertex is taken in the 
end at $t=0$.  Carrying out all the (tedious) calculational steps for the 
right $\Delta^+\Delta^+\gamma$ loop-diagram in Fig.\,6, one arrives at the 
following expression for the radiatively induced magnetic moment:
\begin{eqnarray} F_2^{(\gamma)}(0) &\!\!\! =\!\!\! & {\alpha \kappa^{*2}c
\over 4\pi (3r)^4 } \int_0^\infty\!\!dx\, G^2_\Delta(xM)\bigg\{{(r^2-1)^2
\over x}(1+2r+23r^2+4r^3-27r^4-6r^5+27r^6)\nonumber \\ && +2x(1+r)
(2-r+25r^2 -r^3+r^4-31r^5-r^6+18r^7)\nonumber \\ &&+6x^3(3+r-4r^2-8r^3
-4r^4+6r^5)+2x^5(5r+6r^2-2)+x^7(4r+9r^2-1)\nonumber \\ &&+\sqrt{(r^2-1
+x^2)^2+4x^2}\bigg[{r^2-1\over x}(27r^4+6r^5-27r^6-1-2r-23r^2-4r^3)
\nonumber \\ &&+x(3+27r^2+42r^3+r^4-40r^5-9r^6)+x^3(3-6r-4r^2+4r^3+9r^4)
\nonumber \\ &&+x^5(1-4r-9r^2)+{6r^4x(1+r)(5+9r)\over (1+r)^2+x^2}
\bigg]\bigg\}\,,\end{eqnarray}
where $c=1,0$ takes care that this contribution exists only for the proton.
The dipole times exponential parametrization of $G_\Delta(xM)$ leads to a 
numerical value of $\delta \kappa_p =0.685\cdot 10^{-3}$, which is positive and 
small. The corresponding integrand (without $\alpha/\pi$) is shown by the 
full line in Fig.\,8. One observes a distribution that starts near $x=0$ as 
order $x^3$ and reaches a peak at $x=0.52$.  
\begin{figure}[h!]
\begin{center}
\includegraphics[scale=0.5,clip]{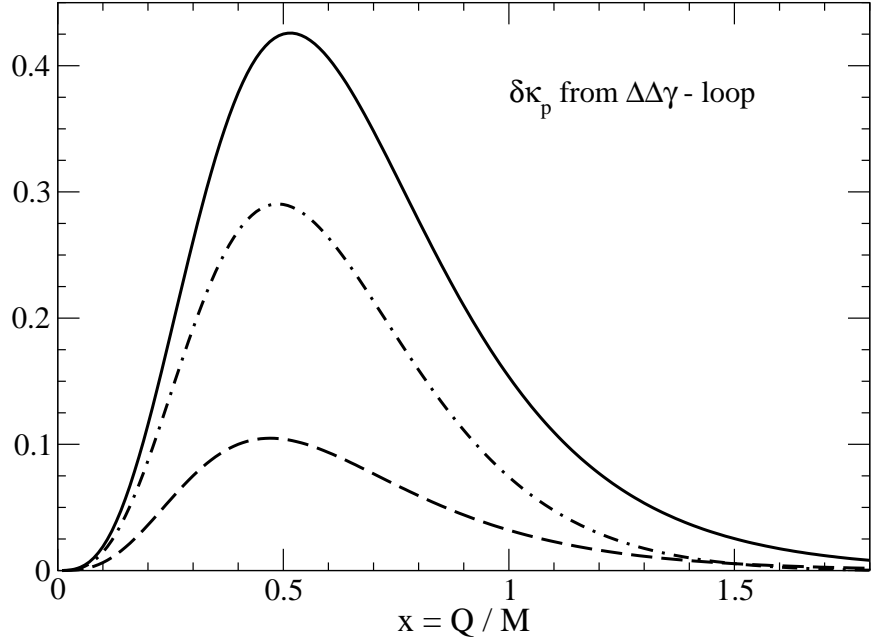}
\end{center}
\vspace{-.6cm}
\caption{Integrand (without factor $\alpha/\pi$) for the radiative 
correction to the proton magnetic moment from the $\Delta^+\Delta^+
\gamma$-loop.}
\end{figure}

After adding up the calculated contributions involving intermediate 
$\Delta(1232)$-isobars, one gets $\delta\kappa_p^{(\Delta)} = -0.87 \cdot 10^{-3}$ 
and $\delta\kappa_n^{(\Delta)} = 1.20 \cdot 10^{-3}$. Together with the purely 
nucleonic loop-contributions from section 3 these yield estimates
for the radiative corrections of:
\begin{equation} 
\delta\kappa_p = -4.29 \cdot 10^{-3}\,, \qquad\qquad \delta\kappa_n = 2.53 \cdot 
10^{-3}\,,\end{equation}
which are still dominated by the $NN\gamma$-loop terms. As measured by
the empirical magnetic moments $1+\kappa_p = 2.793$ and $\kappa_n = -1.913$, 
these values amount to corrections of $-1.5\permil$ and  $-1.3\permil$, 
for the proton and neutron. This is in line with the rough estimate 
$\alpha/2\pi = 1.16\permil$, but opposite in sign.

\section{Alternative forms of spin-3/2 propagator and vertices}   
The Lorentz-covariant treatment of a massive spin-3/2 particle by a 
Rarita-Schwinger spinor-field is affected by various problems concerning the 
presence of spurious spin-1/2 degrees of freedom. For this reason, it has been 
proposed in recent works \cite{ina3,kondr,vanderh} to use for the 
$\Delta$-isobar a modified propagator of the form:
\begin{equation}{i\over 3}\, {\gamma\!\cdot\!P +M_\Delta\over M_\Delta^2
-P^2}\bigg[3 g_{\alpha\beta} -\gamma_\alpha \gamma_\beta-{1\over P^2} \big(
\gamma\!\cdot\!P\gamma_\alpha P_\beta+P_\alpha\gamma_\beta\gamma\!\cdot\!P
\big)\bigg]\,,\end{equation} 
which includes a suitably constructed projector on spin-3/2. The use of such a 
$\Delta$-isobar propagator introduces into our calculation of photon-loop 
diagrams 
a further denominator $P^2$ that depends on the loop-momentum $l$. When 
combining it with the $N\Delta\gamma$-vertex $V_1^{\mu\alpha}$ in eq.(12) one 
gets, as a result of the angular integration $\int_{-1}^1\! dz \sqrt{1-z^2}$, 
weight-functions that have a discontinuity at $x=1$. This feature is 
exemplified in the appendix for the electromagnetic mass shift $\delta M$. We 
consider the appearance of discontinuities as an indication for an inconsistent 
treatment and work with an alternative form of the $\Delta N\gamma$-vertex 
taken from refs.\cite{ina3,kondr,vanderh}:
\begin{equation}V^{\mu \alpha}_2 = {ie\kappa^*\over\sqrt{6M^3M_\Delta }} 
\big(g^{\mu \alpha} \gamma\!\cdot\!P\,\gamma\!\cdot\!l -P^\mu\gamma^\alpha 
\gamma\!\cdot\!l- \gamma^\mu\gamma^\alpha P\!\cdot\!l+ \gamma^\mu 
l^\alpha \gamma\!\cdot\!P\big)\gamma_5\,G_\Delta(\sqrt{-l^2})\,.
\end{equation}
It fulfils the relations $l_\mu V^{\mu\alpha}_2 =0$ (gauge-invariance) and 
$P_\alpha V^{\mu\alpha}_2 =0$ (orthogonality). The prefactor with square-roots of 
masses in the denominator has been chosen such that the residue of the 
spin-independent (real) forward Compton-amplitude $N\gamma\to \Delta \to N\gamma$ 
at the $\Delta(1232)$-pole remains constant. For the first vertex 
$V^{\mu\alpha}_1$ this residue is proportional to $3+r^{-2} = 3.581$, while for
the second vertex $V^{\mu\alpha}_2$ it is (in the same way) proportional to 
$r+3r^{-1}=3.599$, and both numbers agree very well with each other. It should 
be noted that the vertex $V^{\mu \alpha}_2$ in eq.(21)
is just a minimal version with a single coupling constant $\kappa^*$ and 
extensions with further parameters have been proposed in the literature 
\cite{ina3,vanderh}.  
 
We turn now to the reevaluation of the triangular photon-loop diagrams in 
Fig.\,4 with the new $\Delta N\gamma$-vertex $V^{\mu \alpha}_2$. Independent of the
choice of the $\Delta$-isobar propagator (eq.(13) or eq.(20)), this contribution
to the radiatively induced magnetic moment reads:
\begin{eqnarray} F_2^{(\gamma)}(0) &\!\!\! =\!\!\! & {\alpha \kappa^{*2}
\over 54\pi r} \int_0^\infty\!\!dx\, G_\Delta(xM)\bigg\{F_1(xM)
\bigg[-{3+r^2\over x}(r^2-1)^2-12x(1+r^3+x^2r)\nonumber \\ && -4x^5
-{x^2\over r^2-1}\sqrt{4+x^2}(4+12r+10x^2+3x^4)+\sqrt{(r^2-1+x^2)^2+4x^2}
\nonumber \\ && \times \bigg({3+r^2\over x}(r^2-1)+{x \over r^2-1}
\big(9-4r^2+12r^3-r^4+x^2(9+r^2)+3x^4\big)\bigg)\bigg] \nonumber \\ && 
+F_2(xM) \bigg[-{(r^2-1)^2\over 2x}(3+r^2+4r^3)-x(6+2r^2+9r^3+2r^4+3r^5
+2r^6)\nonumber \\ && -6x^3(3+r+2r^2+r^4)+x^5(r-14-6r^2)-2x^7+
{x^2\over r^2-1}\sqrt{4+x^2}\bigg(x^2(r-13)\nonumber \\ &&-8(1+r)-
{7x^4\over 2}\bigg)+{1\over 2}\sqrt{(r^2-1+x^2)^2+4x^2}\bigg({r^2-1
\over x}(3+r^2+4r^3) +{x\over r^2-1} \nonumber \\ && \times \big(9
+14r^3+3r^4+2r^5+4r^6+x^2(9+9r^2-2r^3+8r^4)+x^4(3+4r^2)\big)\bigg)\bigg]
\bigg\}\,,\end{eqnarray}
with the mass ratio $r= M_\Delta/M = 1.312$. One notices that only the part of the 
Rarita-Schwinger propagator proportional to $3g_{\alpha\beta} -\gamma_\alpha 
\gamma_\beta$ has an influence on the result and therefore the additional 
loop-factor $1/P^2$ cannot generate a discontinuity in the weight-functions at 
$x=1$.   

\begin{table}[t!]
\begin{center}
\begin{tabular}{|c|ccc|}
    \hline
form factor & dipole & DRA & Ina \\ \hline  $10^3\cdot\delta\kappa_p$ 
& $-$1.911  & $-$1.900 &$-$1.896  \\ $10^3\cdot\delta\kappa_n$ 
& 1.112  &  1.063 &1.055  \\ 
  \hline
  \end{tabular}
\end{center}
{\it Tab.3: Radiative corrections to the proton and neutron magnetic moments 
arising from $\Delta N\gamma$ triangle-diagrams using with the second vertex.}
\end{table}
 
Table 3 gives the corresponding numerical values of $\delta\kappa_p$ and 
$\delta\kappa_p$ for the three choices (dipole, DRA, Ina) of the nucleon 
vector form factors $F_{1,2}(xM)$. In comparison to the first $\Delta N\gamma
$-vertex, one obtains singificantly reduced  values of $\delta\kappa_p\simeq 
-1.9 \cdot 10^{-3}$ and $\delta\kappa_n\simeq 1.1 \cdot 10^{-3}$. This reduction 
becomes also obvious from the underlying integrands which are shown by the 
dashed lines in Fig.\,5. The differences between the full and dashed lines are 
suggestive of spurious spin-1/2 degrees of freedom in the Rarita-Schwinger 
spinor.

Next we have to reevaluate with the new vertex $V^{\mu \alpha}_2$ the 
wave-function renormalization factor $Z_2^{(\Delta)}$ arising from the 
self-energy subdiagram in Fig.\,6. Independent of the choice of the 
$\Delta$-isobar propagator this quantity is now given by the expression:
\begin{eqnarray}Z_2^{(\Delta)}&\!\!\! =\!\!\! & {\alpha\kappa^{*2}\over 48
\pi r} \int_0^\infty\!\!dx\, G_\Delta^2(xM)\bigg\{{(r^2-1)^2\over x}
(5+2r^2+r^4)+4x(3-r^2+2r^5)+2x^3(3+4r^3)\nonumber \\ &&+4x^5(1+r^2)+3x^7
+\sqrt{(r^2-1+x^2)^2+4x^2}\bigg[{1-r^2\over x}(5+2r^2+r^4)\nonumber \\ 
&& +x (7-4r^2-8r^3+r^4)-x^3(1+r^2)-3x^5+{8r^3x(1+r)\over (1+r)^2+x^2}
\bigg]\bigg\}\,.\end{eqnarray}
The corresponding integrand (without the factor $\alpha/\pi$) is shown by the 
dashed line in Fig.\,7. The apparent enhancement by a factor of about $1.5$ 
translates into more sizeable corrections of opposite sign: $\delta \kappa_p 
=1.732\cdot 10^{-3}$ and $\delta \kappa_n =-1.848\cdot 10^{-3}$.

Finally, we have to reevaluate the $\Delta^+\Delta^+\gamma$ loop-diagram in 
Fig.6 with the new vertex $V^{\mu \alpha}_2$. In this case the result for the 
radiatively induced magnetic moment of the proton depends on the choice the 
$\Delta$-isobar propagator. Using the not spin-3/2 projected version in 
eq.(13), one gets the expression:
\begin{eqnarray} F_2^{(\gamma)}(0) &\!\!\! =\!\!\! & {\alpha \kappa^{*2}c
\over \pi (18r)^2 } \int_0^\infty\!\!dx\, G^2_\Delta(xM)\bigg\{{r\over x}
(r^2-1)^2(23-8r^2+9r^4)+2x^5(18r^3-9r^2+26r-24)\nonumber \\ &&+2x
(8+24r+15r^2-45r^4-12r^5+34r^6)+6x^3(11r^4-8r^2-16)+x^7(27r-16)
\nonumber \\ &&+\sqrt{(r^2-1+x^2)^2+4x^2}\bigg[{r\over x}(1-r^2)
(23-8r^2+9r^4) +x^3(32-25r+2r^2-9r^3)\nonumber \\ && +x(16+25r+46r^2
-20r^3-68r^4+9r^5)+x^5(16-27r)+{6r^3x(1+r)(5+9r)\over (1+r)^2+x^2}
\bigg]\bigg\}\,,\nonumber \\ &&\end{eqnarray} 
which gives the numerical value $\delta \kappa_p =0.415\cdot 10^{-3}$. The 
corresponding integrand is shown by the dashed-dotted line in Fig.\,8. 
On the other hand the spin-3/2 projected version in eq.(20) leads to a somewhat 
different expression: 
\begin{eqnarray} F_2^{(\gamma)}(0) &\!\!\! =\!\!\! & {\alpha \kappa^{*2}c
\over 324\pi r} \int_0^\infty\!\!dx\, G^2_\Delta(xM)\bigg\{{(r^2-1)^2\over
x}(3+8r^2+13r^4)+6x^3(4r+23r^3)\nonumber \\ && +2x(8+9r-45r^3+48r^5
+4r^6)+2x^5(18+21r+14r^2)+23x^7\nonumber \\ &&+\sqrt{(r^2-1+x^2)^2+4x^2}
\bigg[{1-r^2 \over x}(3+8r^2+13r^4)-x^3(13+42r+5r^2)\nonumber \\ && 
+x(13+18r-28r^2-96r^3+5r^4)-23x^5+{6r^2x(1+r)(5+9r)\over (1+r)^2+x^2}
\bigg]\bigg\}\,,\end{eqnarray}
which yields a significantly smaller numerical value of $\delta \kappa_p 
=0.163\cdot 10^{-3}$. This feature is also indicated by the dashed line in Fig.\,8 
for the respective integrand.  It is interesting to note that in the last case 
only the quasi-Dirac part $-i e\, g_{\alpha\beta}\gamma^\mu$ of the $\Delta^+ 
\Delta^+ \gamma$-coupling  has contributed. 
For the spin-3/2 projected vertex and propagator the terms involving 
intermediate $\Delta(1232)$-isobars sum up to $\delta \kappa_p^{(\Delta)} = 
0.00 \cdot 10^{-3}$ and $\delta \kappa_n^{(\Delta)} = -0.79 \cdot  10^{-3}$. 
Together with the purely nucleonic loop-contributions from section\,3 
these yield then estimates for the radiative corrections of:
\begin{equation} 
\delta\kappa_p = -3.42 \cdot 10^{-3}\,, \qquad\qquad \delta\kappa_n = 0.54 \cdot 
10^{-3}\,,\end{equation}
which are (unfortunately) quite different from the first estimate given in 
eq.(19).   

\section{Summary and conclusions}
In this work we have estimated the radiative corrections of order $\alpha/\pi$ 
to the magnetic moments of the proton and neutron. The Schwinger calculation 
for a structureless lepton has been adapted to extended nucleons by including 
phenomenological vector form factors. The resulting expression for 
$F_2^{(\gamma)}(0)$ is infrared-finite and gauge-invariant and yields numerical 
values of $\delta\kappa_p = -3.42\cdot 10^{-3}$ and  $\delta\kappa_n = 1.34
\cdot 10^{-3}$, using updated parametrizations for the nucleon form factors 
$F_{1,2}(Q)$. Moreover, we have also studied the effects of photon-loops 
involving the low-lying $\Delta(1232)$-resonance. For the pertinent calculation 
a Lorentz-covariant treatment of massive spin-3/2 particles is mandatory. The 
two (minimal) versions of the $\Delta N\gamma$-vertex amended by a 
phenomenological transition form factor $G_\Delta(Q)$, that we have examined in 
combination with costumary form of the spin-3/2 propagator, yield radiative 
corrections of $\delta\kappa_p^{(\Delta)}= (-0.9,\, 0.0)\cdot 10^{-3}$ and $\delta
\kappa_n^{(\Delta)} = (1.2,\,-0.8)\cdot 10^{-3}$, respectively. The differences are 
quite substantial and therefore one may hope that an extended form of the 
$\Delta\to N\gamma$ transition amplitude (with further coupling constants and 
off-shell parameters) will lead to radiatively induced nucleon magnetic moments 
that are more consistent for conventional and spin-3/2 projected vertices and 
propagators.  

\section*{Acknowledgements}
I thank A.F. Rey-Fernandez for participation in this project during his bachelor 
thesis. Ina Lorenz has provided valuable information on updated nucleon form 
factor parametrizations and on approaches to the spin-3/2 delta-isobar. 
\section*{Appendix: Electromagnetic mass-shifts}
In this appendix the expressions for electromagnetic mass-shifts of the 
nucleon are given. The calculation of the self-energy subdiagram on the right 
of Fig.\,2 with a $N\gamma$-loop leads to the following gauge-invariant 
(or $\xi$-independent) result:
\begin{eqnarray} \delta M &\!\!\! =\!\!\! & {\alpha M \over 4\pi}
\int_0^\infty\!\!dx\, \bigg\{\big[(2-x^2)\sqrt{4+x^2}+x^3\big] F_1^2(xM) 
+ 3x^2\big(x-\sqrt{4+x^2} \big)\nonumber \\ && \times F_1(xM) F_2(xM)
+ {x^2 \over 8}\big[(x^2-8)\sqrt{4+x^2}+6x-x^3\big] F_2^2(xM) \bigg\}\,,
\end{eqnarray}
with $F_{1,2}(x M)$ the Dirac and Pauli form factors. By evaluating 
this one-parameter integral for the three choices (dipole, DRA, Ina) of 
nucleon form factors, one finds for the electromagnetic proton-neutron mass
difference $\delta M_{p-n}=(0.795,\, 0.757,\,0.782)$\,MeV, respectively.

Next, we consider the self-energy diagram with a $\Delta\gamma$-loop. Using the 
first vertex $V_1^{\mu\alpha}$ in eq.(12) and the conventional Rarita-Schwinger 
propagator in eq.(13), one the obtains the expression:
\begin{eqnarray} \delta M &\!\!\! =\!\!\! & {\alpha M\kappa^{*2}\over 
\pi(12r)^2} \int_0^\infty\!\!dx\, G^2_\Delta(xM)\bigg\{{1+3r^2\over x}
(1-r^2)^3-x^7+6x^3(3-2r^3)\nonumber \\ && -4x^5+4x(1-r^2)(1+r^2+3r^3+r^4)
+\sqrt{(r^2-1+x^2)^2+4x^2}\nonumber \\ && \times \bigg[{1+3r^2
\over x}(r^2-1)^2 +x(3+12r^3+r^4)+x^3(3-r^2)+x^5\bigg]\bigg\}\,, 
\end{eqnarray}
with $G_\Delta(x M)$ the phenomenological $\Delta\to N\gamma$ transition form 
factor. The dipole times exponential function yields a mass-shift of 
$\delta M=0.462$\,MeV, equal for the proton and neutron. When using the second 
vertex $V_2^{\mu\alpha}$ in eq.(21) the result for $\delta M$ is actually 
independent of the choice of the $\Delta$-isobar propagator, and it reads:
\begin{eqnarray} \delta M &\!\!\! =\!\!\! & {\alpha M\kappa^{*2}\over 
144\pi r} \int_0^\infty\!\!dx\, G^2_\Delta(xM)\bigg\{{3+r^2\over x}
(1-r^2)^3+12x(1-r^2)(1+r^3) \nonumber \\ &&  +6x^3(3-4r-2r^3)  
-4x^5(3+r^2)-3x^7+\sqrt{(r^2-1+x^2)^2+4x^2}\nonumber \\ && \times \bigg[
{3+r^2\over x}(r^2-1)^2+x(9-4r^2+12r^3-r^4)+x^3(9+r^2)+3x^5\bigg]
\bigg\}\,. \end{eqnarray}
The corresponding numerical value $\delta M=0.258$\,MeV is reduced by a factor 
$0.56$ compared to the first case. For the sake of completeness we present also 
the expression for $\delta M$ obtained by combining the first vertex 
$V_1^{\mu\alpha}$ with the spin-3/2 projected propagator in eq.(20). It reads:  
\begin{eqnarray} \delta M &\!\!\! =\!\!\! & {\alpha M\kappa^{*2}\over 
144\pi } \int_0^\infty\!\!dx\, G^2_\Delta(xM)\bigg\{\text{sign}(1-x)
{(1+x^2)^4\over r^2x}+{r^2\over x}(8r^2-6-3r^4)\nonumber \\ &&
+4r x(3-3r^2-r^3-3x^2)+{1\over r^2}\sqrt{(r^2-1+x^2)^2+4x^2}\nonumber 
\\ &&\times \bigg[{1+3r^2 \over x}(r^2-1)^2+x(3+12r^3+r^4)+x^3(3-r^2)
+x^5\bigg]\bigg\}\,, \end{eqnarray}
and includes a (peculiar) discontinuous part which stems from the angular 
integral:
\begin{equation} {4\over \pi} \int_{-1}^1\!\!dz\, {x^2 \sqrt{1-z^2} 
\over x^2-1+2i x z}=1-x^2-(1+x^2)\text{sign}(1- x)\,.\end{equation} 
The numerically computed mass-shift $\delta M=0.348$\,MeV lies just in between 
the two previous values. 

\end{document}